%% file: 20icc_fdnn.tex
\documentclass[conference]{IEEEtran}
\IEEEoverridecommandlockouts

\usepackage[dvipsnames]{xcolor}

\usepackage{mathtools}
\usepackage{mathrsfs}
\usepackage{amssymb}
\usepackage{bm}

\usepackage{siunitx}
\DeclareSIUnit{\dBm}{dBm}

\newcommand{\setH}{\mathsf{H}}

\usepackage{array}
\usepackage{makecell}

\usepackage{tikz}
\usepackage{circuitikz}
\usepackage{pgfplots}
\pgfplotsset{compat=1.14}

\usetikzlibrary{calc}
\usetikzlibrary{shapes}
\usetikzlibrary{shapes.misc}
\usetikzlibrary{matrix}
\usetikzlibrary{arrows}
\usetikzlibrary{decorations.markings}
\usetikzlibrary{spy}
\usetikzlibrary{decorations.pathmorphing}
\usetikzlibrary{decorations.markings}
\usetikzlibrary{positioning}
\usetikzlibrary{fit}
\usetikzlibrary{petri}
\usetikzlibrary{chains}
\usetikzlibrary{intersections}
\usetikzlibrary{plotmarks}
\usetikzlibrary{pgfplots.groupplots}

\usepgfplotslibrary{units}
\usepgfplotslibrary{fillbetween}

\pgfplotsset{
    legend image with text/.style={
        legend image code/.code={%
            \node[anchor=center] at (0.3cm,0cm) {#1};
        }
    },
}

\definecolor{darkgreen}{RGB}{0,158,0}
\definecolor{Set1-7-1}{RGB}{178,24,43}
\definecolor{Set1-7-2}{RGB}{166,189,219}
\definecolor{Set1-7-3}{RGB}{214,197,115}
\definecolor{Set1-7-4}{RGB}{39,143,54}
\definecolor{Set1-7-5}{RGB}{122,1,119}
\definecolor{Set1-7-6}{RGB}{244,196,69}
\definecolor{Set1-7-7}{RGB}{178,24,43}

\usepackage{algorithmic}
\usepackage{graphicx}
\usepackage{textcomp}
\usepackage[noadjust]{cite}
\usepackage{threeparttable}
\usepackage{subcaption}
\usepackage{lipsum}
\usepackage{booktabs}
\usepackage{hyperref}
\usepackage{balance}

\def\BibTeX{{\rm B\kern-.05em{\sc i\kern-.025em b}\kern-.08em
    T\kern-.1667em\lower.7ex\hbox{E}\kern-.125emX}}

\newcommand{\figspace}{-0.15cm}

\begin{document}

\title{Identification of Non-Linear RF Systems\\ Using Backpropagation}

\author{\IEEEauthorblockN{Andreas Toftegaard Kristensen\IEEEauthorrefmark{1}, Andreas Burg\IEEEauthorrefmark{1}, and Alexios Balatsoukas-Stimming\IEEEauthorrefmark{2}}
\IEEEauthorblockA{\IEEEauthorrefmark{1}Telecommunication Circuits Laboratory, \'{E}cole polytechnique f\'{e}d\'{e}rale de Lausanne, Switzerland\\
\IEEEauthorrefmark{2}Department of Electrical Engineering, Eindhoven University of Technology, The Netherlands}%
}

\maketitle

\begin{abstract}
In this work, we use deep unfolding to view cascaded non-linear RF systems as model-based neural networks.
This view enables the direct use of a wide range of neural network tools and optimizers to efficiently identify such cascaded models.
We demonstrate the effectiveness of this approach through the example of digital self-interference cancellation in full-duplex communications where an IQ imbalance model and a non-linear PA model are cascaded in series.
For a self-interference cancellation performance of approximately $44.5$~dB, the number of model parameters can be reduced by $74$\% and the number of operations per sample can be reduced by $79$\% compared to an expanded linear-in-parameters polynomial model.
\end{abstract}

\begin{IEEEkeywords}
Backpropagation, memory polynomial, parallel Hammerstein model, Volterra series, SI cancellation.
\end{IEEEkeywords}

\section{Introduction}\label{sec:introduction}

System modeling and identification are some of the most fundamental activities in engineering which allow us to obtain new insights and to make predictions by fitting models to data~\cite{giri2010block}.
While in some cases linear models are sufficient, in general, non-linear models are often required to accurately capture and explain the characteristics of an unknown system.
Such non-linear models have successfully been used for biological systems~\cite{hunter1986identification} and chemical processes~\cite{eskinat1991use}.
One of the most general of these non-linear models is the Volterra series, which models systems with both non-linear and memory effects and is often used as a black-box identification method for non-linear systems~\cite{schetzen1980volterra}. Unfortunately, these generic models are often over-parameterized and have very high complexity.

In communications, the analog front-end of radio frequency (RF) transceivers often introduces severe non-linear distortions.
In particular, power amplifiers (PAs) typically dominate the non-linear behavior of such transceivers and their modeling has been studied extensively (e.g., \cite{kim2001digital}).
This happens because there is a fundamental trade-off between power efficiency and linearity in PAs so that it is desirable to operate them in their non-linear regime.
Digital pre-distortion (DPD) is a linearization technique, where a predistorter applies the inverse of the PA non-linearity to make the overall input-output relationship linear.
If the PA is non-linear, then its inverse is also non-linear and needs to be modeled correspondingly.
Special cases of the Volterra series, such as memory polynomials, which reduce the complexity at the cost of some modeling accuracy, are more commonly used in practice for DPD~\cite{morgan2006generalized}.

Non-linear RF system identification becomes more challenging when the effect of the remaining components (i.e., other than the PA) in the transceiver chain cannot be ignored.
This is, for example, the case for in-band full-duplex (FD) communications where an extremely accurate model of the self-interference (SI) needs to be constructed to achieve a high level of cancellation~\cite{Bharadia2013}.
Only modeling PA non-linearities is often not sufficient to cancel the SI to the level of the receiver noise floor and the effects of other transceiver blocks, such as digital-to-analog converter (DAC) non-linearities~\cite{Balatsoukas2015} and IQ imbalance~\cite{Korpi2014}, also need to be modeled.
In principle, a general black-box model, such as the Volterra series, can be used to capture the combined effect of multiple sources of non-linearity.
However, the complexity of this approach can be very high and, thus, prohibitive for practical applications.

\subsubsection*{Contribution}
In this work, we examine non-linear RF system identification through the lens of neural networks (NNs) by applying the concept of \emph{deep unfolding}~\cite{hershey2014deep,balatsoukas2019deep} to a cascade of non-linear systems and by using backpropagation to tune the (complex-valued) model parameters.
This approach is similar to~\cite{yu2014identification}, where a Hammerstein-type NN is derived.
However, our models are complex-valued, we cascade the polynomial model with the additional non-linearity of an IQ-mixer and we use a standard NN framework which allows us to employ the full toolbox available for researchers to train such non-linear models.
The presented approach also allows us to capture unknown effects by including conventional NNs (e.g., feed-forward NNs) as part of the cascade of non-linear models in a straightforward fashion.
In the general case of multiple cascaded non-linear models, the deep unfolding approach significantly reduces the number of parameters to be estimated and the number of floating-point operations per produced output sample (FLOPs) with respect to a linear-in-parameters non-linear model.
To demonstrate the effectiveness of the deep unfolding approach, we show the application of these non-linear models for digital SI cancellation in FD communications where both IQ imbalance and PA non-linearities have to be modeled.
To the best of our knowledge, this is also the first work that uses model-based NNs for SI cancellation.

\section{Volterra-Series-Based Non-Linear Models}\label{sec:background}

The Volterra series is a very general (discrete) non-linear model with a memory of $M$ samples, which consists of a sum of multidimensional convolutions~\cite{schetzen1980volterra}:
\begin{equation}\label{eq:volterra}
	y[n] = \sum_{p=1}^{P} \sum_{m_1=0}^{M-1} \cdots \sum_{m_P=0}^{M-1} h_p[m_1, \dots, m_p] \prod_{l=1}^{p} x[n-m_l] ,
\end{equation}
where $P$ is the maximum non-linearity order, $h_p[m_1, \dots, m_p]$ represent the \emph{Volterra kernels} (i.e., parameters to be fitted), $x$ represents the input signal, and $y$ represents the output signal.
The number of terms in~\eqref{eq:volterra} is in the order of $M^P$, which motivates the use of simpler models in practice.

\subsection{Simplified Non-Linear Models}
\begin{figure}[t]
	\centering
	\begin{subfigure}{0.9\linewidth}
		\centering
		\scalebox{0.75}{\input{fig/wiener.tikz}}
		\caption{Wiener model.}
		\label{fig:wiener}
	\end{subfigure}
	\vskip\baselineskip
	\begin{subfigure}{0.9\linewidth}
		\centering
		\scalebox{0.75}{\input{fig/hammerstein.tikz}}
		\caption{Hammerstein model.}
		\label{fig:hammerstein}
	\end{subfigure}
	\vskip\baselineskip
	\begin{subfigure}{0.9\linewidth}
		\centering
		\scalebox{0.75}{\input{fig/hammerstein_wiener.tikz}}
		\caption{Hammerstein-Wiener model.}
		\label{fig:hammerstein_wiener}
	\end{subfigure}
	\vskip\baselineskip
	\begin{subfigure}{0.9\linewidth}
		\centering
		\scalebox{0.75}{\input{fig/hammerstein_parallel.tikz}}
		\caption{Parallel Hammerstein model.}
		\label{fig:hammerstein_parallel}
	\end{subfigure}
\caption{Common non-linear models, with $H(\cdot)$ blocks representing LTI blocks, i.e., filters, and $f(\cdot)$ blocks representing static non-linear functions.}
\label{fig:models}
\vspace{\figspace}
\end{figure}
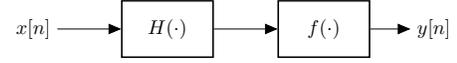
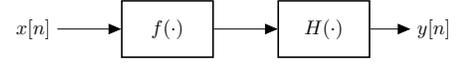
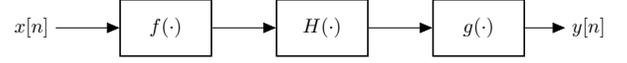
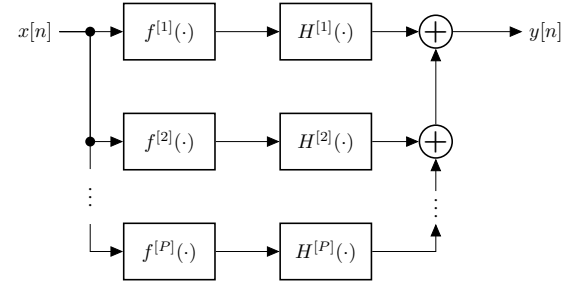

The \emph{Wiener model}, shown in Fig.~\ref{fig:wiener}, is a simplification of the Volterra series, which consists of a linear filter followed by a static non-linearity:
\begin{equation}\label{eq:wiener}
	y[n] = \left(\sum_{m=0}^{M-1} h[m] x[n-m] \right)^P .
\end{equation}
This is equivalent to setting $h_p[m_1, \dots, m_p] = h[m_1] \cdots h[m_p]$ in~\eqref{eq:volterra} and only considering the term for $p=P$.
However, the output now depends \emph{non-linearly} on the coefficients $h[m]$, thus making their estimation difficult.

In the \emph{Hammerstein model}, shown in Fig.~\ref{fig:hammerstein}, a static non-linearity is followed by a linear filter:
\begin{equation}\label{eq:hammerstein}
	y[n] =   \sum_{m=0}^{M-1} h[m]  \left(x[n-m]\right)^P.
\end{equation}
The Hammerstein model has the desirable property that it is linear in the parameters $h[m]$, thus enabling the use of linear regression for parameter estimation.

In order to express more complex non-linearities, the Wiener and Hammerstein models can be concatenated  \emph{in series} and \emph{parallel}. One example of a concatenation in series is the Hammerstein-Wiener model, shown in Fig.~\ref{fig:hammerstein_wiener}, given by:
\begin{equation}\label{eq:hammerstein_wiener}
	y[n] = \left(\sum_{m=0}^{M-1} h_P[m]  \left(x[n-m]\right)^P \right)^Q ,
\end{equation}
where $Q$ represents the power for the non-linearity $g(\cdot)$ in Fig.~\ref{fig:hammerstein_wiener}. An example of a parallel Hammerstein model is shown in Fig.~\ref{fig:hammerstein_parallel}, where each non-linearity creates a branch and the outputs of all branches are added together:
\begin{equation}\label{eq:hammerstein_parallel}
	y[n] =   \sum_{m=0}^{M-1} \sum_{p=1}^{P} h_p[m]  \left(x[n-m]\right)^p.
\end{equation}

The above models are designed to be generic and not matched to the characteristics of a specific problem. However, in communications applications, non-linear models are typically based on a \emph{complex-valued} baseband signal.
In this case, the parallel Hammerstein model in~\eqref{eq:hammerstein_parallel} is typically re-written as a \emph{memory polynomial}~\cite{kim2001digital, morgan2006generalized}:
\begin{equation}\label{eq:memory_polynomial}
	y[n] = \sum_{m=0}^{M-1} \sum_{p=1}^{P} h_p[m] x[n-m] \left| x[n-m] \right|^{p-1} .
\end{equation}

\subsection{Parameter Estimation}

If the output is linear in terms of the parameters, as is the case for the Hammerstein models and the memory polynomial, then a closed-form solution can be obtained using linear least-squares (LS) as follows.
First, we collect the coefficients in a vector $\mathbf{h}$ of size $K \times 1$, where $K$ is the total number of parameters to be estimated.
The inputs are arranged into a matrix $\mathbf{X}$ of shape $N \times K$ so that the predictions can be written as $\mathbf{{y}} = \mathbf{X} \mathbf{h}$, where $\mathbf{{y}}$ is an $M \times 1$ vector that is an estimate of the actual output $\mathbf{t}$.
The parameters $\mathbf{\hat{h}}$ minimizing $|| \mathbf{t} - \mathbf{{y}} ||^2$ are then obtained as:
\begin{equation}
	\mathbf{\hat{h}} = \left(\mathbf{X}^\setH \mathbf{X} \right)^{-1} \mathbf{X}^\setH \mathbf{y} .
\end{equation}

Iterative methods based on gradient descent can also be applied as an alternative to LS in order to avoid the high-complexity matrix inversion operation. Another advantage of gradient-descent-based algorithms is that they can be used in conjunction with gradient backpropagation to identify models that are not  linear in their parameters, such as the Wiener model and serial concatenations of non-linear models.

\section{Learning the Memory Polynomial}\label{sec:learning}

In this section, we explain how a gradient-descent-based algorithm can be derived using Wirtinger calculus through the simple example of the commonly used complex-valued memory polynomial in~\eqref{eq:memory_polynomial}.
We then also consider a more challenging serially concatenated non-linear model that models the effect of both the IQ mixer and the PA in an RF transmitter.

\subsection{Gradient Descent Using Wirtinger Calculus}

The parameters of a non-linear system are typically chosen so that a real-valued scalar cost function of the complex-valued output is minimized.
However, a real-valued function of complex variables is generally not complex analytic and therefore not complex differentiable.
To perform gradient descent on functions that are not complex differentiable, we use Wirtinger calculus~\cite{wirtinger1927formalen}.
Specifically, let $z\in \mathbb{C}$ and $f(z) \in \mathbb{R}$, and let $\overline{z}$ denote the complex conjugate of $z$.
The direction of steepest ascent for $f$ is given by the derivative of $f$ with respect to $\overline{z}$, i.e., $\frac{\partial f(z)}{\partial \overline{z}}$.
Using Wirtinger calculus, the derivative $\frac{\partial f(z)}{\partial \overline{z}}$ can be calculated by re-writing $f(z)$ as a bi-variate function of $z$ and $\overline{z}$, $f(z, \overline{z})$, and then treating $\overline{z}$ as the variable and $z$ as a constant.
The derivative chain rule for a composition of $f$ with another function $g$, i.e., $f(g(z))$, is given by:
\begin{align}
	\frac{\partial f(g(z))}{\partial z} &= \frac{\partial f}{\partial g} \frac{\partial g}{\partial z} + \frac{\partial f}{\partial \overline{g}} \frac{\partial \overline{g}}{\partial z} \label{eq:chain} , \\
	\frac{\partial f(g(z))}{\partial \overline{z}} &= \frac{\partial f}{\partial g} \frac{\partial g}{\partial \overline{z}} + \frac{\partial f}{\partial \overline{g}} \frac{\partial \overline{g}}{\partial \overline{z}} . \label{eq:chain_conj}
\end{align}
With these rules and definitions, any gradient-descent-based algorithm minimizing a real-valued cost function $C(\cdot)$ of complex-valued variables to optimize a set of parameters $\mathbf{h} \in \mathbb{C}^{K}$ is based on the following update scheme:
\begin{equation}\label{eq:grad}
	\mathbf{\hat{h}} = \mathbf{h} - \lambda \nabla_\mathbf{h} C, \quad \text{with} \, \nabla_\mathbf{h} C \triangleq \left[ \frac{\partial C}{\partial \overline{h}_1}, \dots, \frac{\partial C}{\partial \overline{h}_{K}} \right]^\top ,
\end{equation}
where $\lambda$ is called the \emph{step size} and $\nabla_\mathbf{h} C$ can be determined by repeatedly applying the chain rule in~\eqref{eq:chain_conj}.

\subsection{Learning the Memory Polynomial}\label{sec:lmp}

We now explain how the parameters of~\eqref{eq:memory_polynomial} can be learned using gradient descent.
We use the complex-valued mean-squared error (MSE) cost function:
\begin{equation}
	 \mathrm{C}(t[n],y[n]) = \frac{1}{2} ({t}[n] - y[n])\overline{({t}[n] - y[n])},
\end{equation}
which we hereafter denote by $C$ to simplify the notation.
In order to derive the gradient of the cost function $C$ with respect to the parameters $\mathbf{h}$, the first step is to derive the gradient of $C$ with respect to the output $y[n]$:
\begin{align}\label{eq:grad_y}
	\nabla_{y[n]} C &= \frac{\partial}{\partial \overline{y}[n]} \frac{1}{2}  (t[n] - y[n])\overline{(t[n] - y[n])} = \frac{1}{2} \left( y[n]-t[n]  \right).
\end{align}
Next, we can obtain the gradient with respect to any parameter $h_q[l]$, by applying the chain rule in~\eqref{eq:chain_conj}:
\begin{equation}
	\nabla_{h_q[l]} C = \frac{\partial C}{\partial y[n]} \frac{\partial y[n]}{\partial \overline{h}_q[l]} + \frac{\partial C}{\partial \overline{y}[n]} \frac{\partial \overline{y}[n]}{\partial \overline{h}_q[l]} .
\end{equation}
Since $y[n]$ is independent of $\overline{h}_q[l]$, we have that $\frac{\partial y[n]}{\partial \overline{h}_q[l]}=0$. Moreover, the derivative $ \frac{\partial \overline{y}[n]}{\partial \overline{h}_q[l]}$ is given by:
\begin{align}
	\frac{\partial \overline{y}[n]}{\partial \overline{h}_q[l]} &= \frac{\partial}{\partial \overline{h}_q[l]}  \sum_{m=0}^{M-1} \sum_{p=1}^{P} \overline{h}_p[m] \overline{x}[n-m] \left| x[n-m] \right|^{p-1} \notag \\
&= \overline{x}[n-l] \left| x[n-l] \right|^{q-1} ,
\end{align}
so that the gradient of $C$ with respect to $h_q[l]$ is:
\begin{align}\label{eq:grad_memory_polynomial}
	\nabla_{h_q[l]}C &=  \frac{1}{2} \left( y[n]-t[n]  \right) \overline{x}[n-l] \left| x[n-l] \right|^{q-1} .
\end{align}
Using~\eqref{eq:grad_memory_polynomial} with~\eqref{eq:grad} is sufficient for updating the parameters of~\eqref{eq:memory_polynomial} with gradient descent.
Note that this algorithm for updating the filter weights $\mathbf{h}$ is, in fact, the well-known complex-valued LMS algorithm for linear system identification~\cite{widrow1975complex}.

\begin{figure}[t]
	\centering
	\scalebox{0.9}{\input{fig/system.tikz}}
	\caption{Simplified two-antenna wireless transceiver block diagram. The channel $h_{\text{SI}}$ models the self-interference channel discussed in Section~\ref{sec:results}.}
	\label{fig:fd_system}
	\vspace{\figspace}
\end{figure}
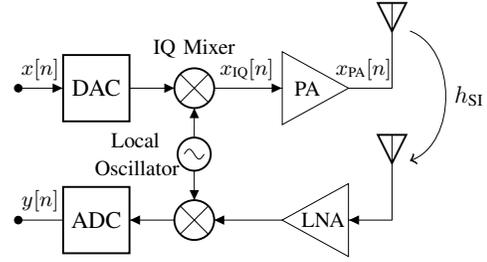

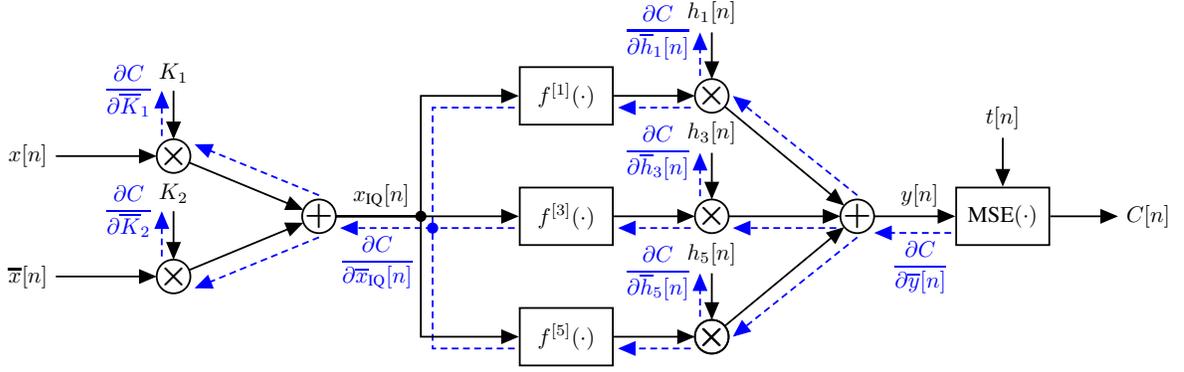
\begin{figure*}[t]
	\centering
	\scalebox{0.85}{\input{fig/forward_backward.tikz}}
	\caption{Model-based NN obtained by unfolding \eqref{eq:model_based_nn} for $M=1$ and $P=5$, with $f^{[P]}(z)=z|z|^{P-1}$ as the non-linearity and the MSE cost function applied at the end.
	Black arrows show forward propagation and blue dashed arrows show backpropagation.}
	\label{fig:forward_backward}
	\vspace{\figspace}
\end{figure*}

\section{Learning Concatenated Non-Linear Systems}

While the memory polynomial in~\eqref{eq:memory_polynomial} can accurately model the PA in a transmitter chain, other components such as the IQ mixer, shown in Fig.~\ref{fig:fd_system}, can also introduce significant non-linear effects.
While the combined effect of the PA and the IQ mixer could be modeled using a black-box approach (e.g., using a Volterra series), it is more efficient to model each non-linearity separately with an appropriate model and to concatenate the individual non-linearities serially, thus exploiting expert knowledge about the block structure of the transceiver which simplifies the overall model.

\subsection{System with IQ Imbalance and PA Non-Linearities}
The IQ imbalance non-linearity can be modeled as~\cite{Korpi2017}:
\begin{equation}\label{eq:iq}
	x_{\text{IQ}}[n] = K_1x[n] + K_2\overline{x}[n],
\end{equation}
where $x[n]$ is the complex-valued baseband equivalent of the IQ mixer input signal, $K_1, K_2 \in \mathbb{C}$ are complex-valued parameters, and typically $|K_1| \gg |K_2|$.
The output of the IQ mixer is amplified by the PA, which introduces additional non-linearities that can be modeled using a memory polynomial:
\begin{equation}\label{eq:pa}
	x_{\text{PA}}[n] = \sum_{m=0}^{M-1} \sum _{\substack{p=1,\\p \text{ odd}}}^P h_{\text{PA},p}[m]x_{\text{IQ}}[n-m]|x_{\text{IQ}}[n-m]|^{p-1},
\end{equation}
where $h_{\text{PA},p}$ is the impulse response for the $p$-th order PA non-linearity and $M$ is the memory length of the PA.
If we substitute~\eqref{eq:iq} into~\eqref{eq:pa} and expand the powers of $x_{\text{IQ}}$, we obtain the following expression for the output of the PA~\cite{Korpi2017}:
\begin{equation}\label{eq:final}
	x_{\text{PA}}[n] = \sum _{\substack{p=1,\\p \text{ odd}}}^P \sum_{q=0}^p\sum_{m=0}^{M-1} g_{p,q}[m] x[n-m]^{q}\overline{x}[n-m]^{p-q} ,
\end{equation}
where $g_{p,q}[m]$ is a channel containing the combined effects of $K_1$, $K_2$, and $h_{\text{PA},p}$.
We note that only odd values of $p$ are considered in~\eqref{eq:final} because the harmonics for even values of $p$ typically lie out of band and are filtered out by a passband filter before the antenna, which is not shown in Fig.~\ref{fig:fd_system} for simplicity.
The expression in~\eqref{eq:final} is linear in its parameters, which allows us to apply the LMS algorithm to efficiently estimate them. However, with this approach, the number of parameters in~\eqref{eq:final} is $N_p = \frac{M}{4}(P+1)(P+3)$ which scales quadratically with $P$ and the model is typically overparameterized~\cite{Korpi2017}. This problem becomes progressively more severe as more non-linearities are concatenated, e.g., if we also want to model the DAC non-linearities.

Instead of using~\eqref{eq:final} with LMS-based estimation, it is possible to substitute~\eqref{eq:iq} in~\eqref{eq:pa} and to directly learn the following expanded model:
\begin{align}\label{eq:model_based_nn}
x_{\text{PA}}[n] = \sum _{\substack{p=1,\\p \text{ odd}}}^P \sum_{m=0}^{M-1} & h_{\text{PA},p}({K_1}x[n-m] + {K_2}\overline{x}[n-m]) \nonumber \\ & \times |({K_1}x[n-m] + {K_2}\overline{x}[n-m])|^{p-1} .
\end{align}
This can be achieved by using deep unfolding with gradient backpropagation and gradient descent as shown in Fig.~\ref{fig:forward_backward}, which significantly reduces the total number of parameters because the individual physical parameters are learned directly, thus avoiding over-parameterization. Specifically, it can be seen that the number of parameters in \eqref{eq:model_based_nn} is $N_p = \frac{1}{2}(P+1)M+2$, which only scales linearly with $P$.

\subsection{Learning IQ Imbalance and PA Non-Linearities}

The addition of the IQ imbalance non-linearity does not affect the PA non-linearity, for which we have already derived the necessary gradients in Section~\ref{sec:lmp}. Specifically, we can re-use~\eqref{eq:grad_y} with $x_{\text{IQ}}$ replacing $x$ and we only have to determine two additional gradients with respect to the parameters $K_1$ and $K_2$.
To this end, we can obtain the gradient with respect to $x_{\text{IQ}}[r]$, with $r=n-l$, by applying the chain rule:
\begin{equation}
	\nabla_{x_{\text{IQ}}[r]} C = \frac{\partial C}{\partial {y}[n]} \frac{\partial {y}[n]}{\partial \overline{x}_{\text{IQ}}[r]} + \frac{\partial C}{\partial \overline{y}[n]} \frac{\partial \overline{y}[n]}{\partial \overline{x}_{\text{IQ}}[r]}.
\end{equation}
We have $ \frac{\partial C}{\partial y[n]} = \overline{\left( \frac{\partial C}{\partial \overline{y}[n]} \right)}$ as $C \in \mathbb{R}$. We also have:
\begin{align}
	\frac{\partial {y}[n]}{\partial \overline{x}_{\text{IQ}}[r]} &= \frac{\partial}{\partial  \overline{x}_{\text{IQ}}[r]}  \sum_{m=0}^{M-1} \sum_{\substack{p=1,\\p \text{ odd}}}^{P} {h}_p[m] {x}_{\text{IQ}}[n-m] \left| x_{\text{IQ}}[n-m] \right|^{p-1} \notag \\
&= \frac{1}{2} \sum_{\substack{p=1,\\p \text{ odd}}}^{P} {h}_p[l] |{x}_{\text{IQ}}[n-l]|^{p-1} \frac{{x}_{\text{IQ}}[n-l]}{\overline{x}_{\text{IQ}}[n-l]} (p-1) ,
\end{align}
and, similarly:
\begin{align}
\frac{\partial \overline{y}[n]}{\partial \overline{x}_{\text{IQ}}[r]}
&= \frac{\partial}{\partial  \overline{x}_{\text{IQ}}[r]}  \sum_{m=0}^{M-1} \sum_{\substack{p=1,\\p \text{ odd}}}^{P} \overline{h}_p[m] \overline{x}_{\text{IQ}}[n-m] \left| x_{\text{IQ}}[n-m] \right|^{p-1} \notag \\
&= \frac{1}{2} \sum_{\substack{p=1,\\p \text{ odd}}}^{P} \overline{h}_p[l] |{x}_{\text{IQ}}[n-l]|^{{p-1}} (p+1) .
\end{align}
Finally, we can obtain the gradients with respect to the parameters $K_1$ and $K_2$ as:
\begin{align}
	\nabla_{K_1} C &= \frac{\partial C}{\partial y[n]} \frac{\partial y[n]}{\partial \overline{K}_1} + \frac{\partial C}{\partial \overline{y}[n] } \frac{\partial \overline{y}[n] }{\partial \overline{K}_1} \notag \\
	&=  \sum_{m=0}^{M-1}  \frac{\partial \overline{x}_{\text{IQ}}[n-m] }{\partial \overline{K}_1} \nabla_{x_{\text{IQ}}[n-m]} C \notag \\
	&=  \sum_{m=0}^{M-1}   \overline{x}[n-m] \nabla_{x_{\text{IQ}}[n-m]} C \label{eq:grad_k1} ,
\end{align}
and, similarly:
\begin{align}\label{eq:grad_k2}
	\nabla_{K_2} C = \sum_{m=0}^{M-1} x[n-m] \nabla_{x_{\text{IQ}}[n-m]}  C .
\end{align}
All gradient backpropagation steps are illustrated with blue color in Fig.~\ref{fig:forward_backward} for a simple case where $M=1$ and $P=5$.
Using~\eqref{eq:grad_memory_polynomial}, \eqref{eq:grad_k1}-\eqref{eq:grad_k2}, and~\eqref{eq:grad} is sufficient to update all parameters of~\eqref{eq:model_based_nn} with gradient-descent-based algorithms.

\subsection{Discussion}
The approach described in the previous section is a simple non-linear optimization method that can be applied to any composition of functions and, hence, any cascade of transceiver RF block models. However, its main shortcoming is that it can get stuck in local minima of the cost function.
In essence, we are shifting the main problem of identifying cascaded non-linear systems from overparameterization and high complexity (cf. the discussion after \eqref{eq:final}) to a more difficult training process.
This is exactly the problem that NNs also have. Although the cost function landscape may look different, the various optimizers that have been used successfully for NNs can also be used directly with unfolded cascaded non-linear systems.
Moreover, gradient computation seems like a daunting task, especially if many non-linear systems are cascaded.
Fortunately, NN tools that support complex-valued arithmetics and automatic differentiation, such as TensorFlow, make this task much easier since each non-linearity can simply be described as a NN-like layer.

\section{Case Study: Self-Interference Cancellation for In-Band Full-Duplex Communications}\label{sec:results}

In this section, we present training and inference results for the non-linear SI cancellation in FD radios as an example of the above-described methodology.
Consider the basic block diagram of a FD transceiver as shown in Fig.~\ref{fig:fd_system}.
If no signal-of-interest from a remote node is present, the received signal $y[n]$ is the SI signal, which is a non-linear function of the transmitted baseband signal.
The goal of the non-linear SI cancellation is to compute an estimate of the SI signal $y[n]$, denoted by $\hat{y}[n]$, based on the transmitted baseband signals $x[n]$.
The estimated SI signal $\hat{y}[n]$ is then subtracted from $y[n]$, so that the residual SI signal is as close to zero as possible.
The SI cancellation performance is therefore evaluated using the following performance measure over a window of $N$ samples:
\begin{align}
	C_{\text{dB}} & = 10\log_{10}\left( \frac{\sum _{n=0}^{N-1}|y[n]|^2}{\sum _{n=0}^{N-1}|y[n] - \hat{y}[n]|^2}\right) . \label{eq:perf}
\end{align}
We use the term \emph{widely-linear memory polynomial} (WLMP) to refer to the model in \eqref{eq:final} with LS parameter estimation and the term \emph{model-based NN} to refer to our model in \eqref{eq:model_based_nn} where deep unfolding and backpropagation are used to estimate the parameters.
The goal of this section is to compare the WLMP with the model-based NN in terms of the achieved SI cancellation performance and in terms of their complexity.
\begin{table}[t!]
	\centering
	\caption{Average SI cancellation across 20 initializations with 50 epochs for the model-based NN (MB-NN) with and without IQ imbalance, as well as the LS-fitted WLMP model.}
	\label{tab:poly_model_comparison}
	\begin{tabular}{lrrr}
		\toprule
		$P$	& \multicolumn{1}{c}{MB-NN w/o IQ imb.}	& \multicolumn{1}{c}{MB-NN w. IQ imb.} 			& \multicolumn{1}{c}{WLMP} \\
		\midrule
		3 		& $42.4\pm 0.03$ \SI{}{\decibel} 		& $43.2\pm 0.06$ \SI{}{\decibel} 	& \SI{43.7}{\decibel} \\
		5 		& $42.0\pm 0.25$ \SI{}{\decibel}		& $44.4\pm 0.06$ \SI{}{\decibel} 	& \SI{44.5}{\decibel} \\
		7 		& $28.8\pm 3.05$ \SI{}{\decibel}		& $44.6\pm 0.11$ \SI{}{\decibel} 	& \SI{44.8}{\decibel}\\
		9 		& $2.9\pm 3.20$ \SI{}{\decibel}			& $44.5\pm 0.32$ \SI{}{\decibel} 	& \SI{44.5}{\decibel}\\
		\bottomrule
	\end{tabular}
	\vspace{\figspace}
\end{table}
\subsection{Dataset \& Experimental Setup}
All considered models are fitted on an SI dataset consisting of QPSK-modulated OFDM signals with a bandwidth of \SI{10}{\mega\hertz} and $1024$ carriers, sampled at \SI{20}{\mega\hertz}.
The transmitted OFDM frame consists of $\sim$\SI{20\,000}{} baseband samples, with \SI{90}{\percent} used for training and the remaining \SI{10}{\percent} used for testing.
An average transmit power of \SI{10}{\dBm} is used and the two-antenna setup provides a passive suppression of \SI{53}{\decibel}.
Active analog cancellation is not used as the achieved passive suppression is sufficient.
The testbed and the dataset are described in more detail in~\cite{Balatsoukas2015} and \cite{Kristensen2019}, respectively.

Similarly to the work of~\cite{Kristensen2019}, which uses the same dataset, both the WLMP and the model-based NN are considered for $P \in \{3, 5, 7, 9\}$ and $M=13$. We use LS estimation for the WLMP\footnote{We note that this is a conservative choice in the sense that we allow the WLMP to use a one-shot training method while the model-based NN uses a mini-batch gradient descent training method, which generally has worse performance.} and the FTRL optimizer for the model-based NN with the default values for all parameters except for the batch size and the learning rate, which are selected using a random hyperparameter search.
Finally, as is common practice when training NNs, we normalize the input and output training samples so that they have unit variance (i.e., the variance of the real part and the variance of the imaginary part are both equal to $0.5$) and zero mean.
We note that the code for training our models as well as the employed dataset are publicly available at \url{https://github.com/A-T-Kristensen/rf\_unfolding}.

\subsection{Results}

\subsubsection{SI Cancellation Performance}
Table~\ref{tab:poly_model_comparison} shows the performance of the LS-fitted WLMP model and the average SI cancellation achieved by the model-based NN over 20 different initializations after training for 50 epochs.
The model-based NN is initialized in a similar way as~\cite{he2015delving}, such that the initial output variance is one and the mean is zero.
We show results for the model-based NN with and without the IQ imbalance layer to highlight that IQ imbalance is indeed a significant non-linearity in this dataset.
We observe that the model-based NN and the LS-fitted WLMP model achieve almost identical SI cancellation performance for all values of $P$.
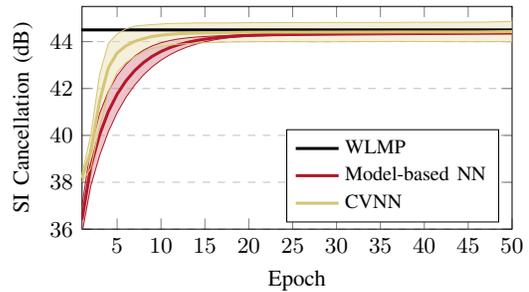
\begin{figure}[t!]
	\centering
	\scalebox{0.95}{\input{fig/train_plot_hammerstein.tikz}}
	\caption{Average cancellation for 20 initializations $\pm 1$ standard deviation for the model-based NN with $P=5$ and the CVNN with 6 hidden neurons.}
	\label{fig:train_all}
	\vspace{\figspace}
\end{figure}
\subsubsection{Training}
Fig.~\ref{fig:train_all} shows the training curve for the model-based NN for $P=5$ and a complex-valued NN (CVNN) from~\cite{Kristensen2019} with 6 neurons in the hidden layer for $20$ different random initializations.
For the model-based NN, we observe that the variation across the different initializations reduces significantly after a certain number of epochs and at each epoch, the average cancellation always either improves or remains relatively stationary.
Moreover, in our experiments, we observed that the training of the model-based NN converges to practically the same values for the parameters independently of the initialization.
We observe that the CVNN converges faster than the model-based NN, however, it has a higher variation across different initializations.
The FTRL optimizer used for the model-based NN is similar to online gradient descent, except that a per-parameter learning rate schedule is used instead of a global one.
The per-parameter learning rate schedule makes the performance over different initializations much more stable for the model-based NN, compared to experiments where we used mini-batch gradient descent with a global learning rate schedule.
As such, the model-based NNs show very stable training performance when using the FTRL optimizer.
\looseness=-1

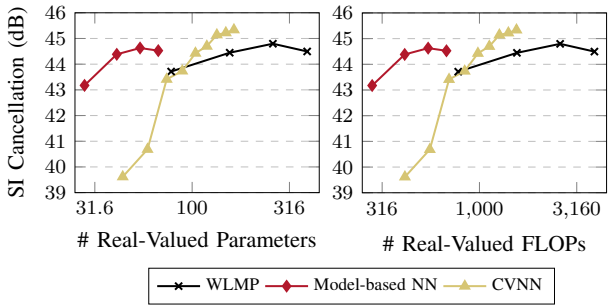
\begin{figure}[t!]
	\centering
	\scalebox{0.95}{\input{fig/complexity_group_plot.tikz}}
	\caption{Test set performance for all models as a function of their memory and computational complexity.}
	\label{fig:perf_compare}
	\vspace{\figspace}
\end{figure}
\subsubsection{Complexity}
Fig.~\ref{fig:perf_compare} shows the SI cancellation performance of the WLMP, the model-based NN, as well as the black-box CVNN from~\cite{Kristensen2019} as a function of the number of real-valued FLOPs and the number of real-valued parameters.
The number of real-valued FLOPs and the number of complex-valued parameters are a proxy for the computational complexity and the memory requirements of the various SI cancelers, respectively~\cite{Kristensen2019}.
We observe that the model-based NN achieves significantly better performance for the same complexity than the WLMP. We also observe that the CVNN achieves a better maximum performance than both the model-based NN and the WLMP. This motivates the concatenation of the model-based NN with a complex-valued NN layer to model still-unknown effects, which can be achieved very easily with the deep unfolding approach.

For $P=5$, the WLMP has $N_p = 156$ complex-valued parameters and requires $1558$ FLOPs, while the model-based NN only has $N_p = 41$ complex-valued parameters and requires $331$ FLOPs.
A CVNN from~\cite{Kristensen2019} that achieves an SI cancellation performance of $44.4$~dB, similar to the WLMP and model-based NN for $P=5$, has $N_w= 119$ complex-valued weights and requires $1124$ FLOPs.
As such, for almost identical SI cancellation performance, the model-based NN has $74$\% and $66$\% fewer parameters than the WLMP and the CVNN, respectively, while also requiring $79$\% and $71$\% fewer FLOPs than the WLMP and the CVNN, respectively.

\section{Conclusion}\label{sec:conclusion}

In this paper, we proposed non-linear RF system identification by applying the concept of deep unfolding to a cascade of different blocks of non-linear systems. The corresponding complex-valued parameters can then be tuned by using backpropagation.
We used SI cancellation in FD radios as an example to show that, by preserving the original model structure, the model-based NNs can significantly and consistently reduce the number of parameters and the FLOPs compared to a standard LS-fitted widely linear memory polynomial as well as compared to a black-box complex-valued NN.


\bstctlcite{IEEEexample:BSTcontrol}
\bibliographystyle{IEEEtran}
\bibliography{IEEEabrv,bibliography}

\end{document}

%% file: fig/wiener.tikz
\def\horizontalsep{3cm}
\def\vertsep{7cm}

\def\filterheight{1cm}
\def\filterwidth{1.618cm}

\begin{tikzpicture}[-triangle 45, scale=0.65]
  \tikzstyle{dspfilter}=[rectangle, draw=black, minimum width=\filterwidth, minimum height=\filterheight, inner sep=2pt, thick]
	
	\node[] (input) at (0,0) {$x[n]$};
	\node[dspfilter] (filter) at ($(input.east)+ (\horizontalsep,0)$) {$H(\cdot)$};  
	\node[dspfilter] (function) at ($(filter.east)+ (\horizontalsep,0)$) {$f(\cdot)$};    
	\node[] (output) at ($(function)+ (\horizontalsep,0)$) {$y[n]$};
	
  
	\draw[-triangle 45] (input) -- (filter);
	\draw[] (filter) -- (function);
	\draw[] (function) -- (output);	
	
\end{tikzpicture}

%% file: fig/hammerstein.tikz
\def\horizontalsep{3cm}
\def\vertsep{7cm}

\def\filterheight{1cm}
\def\filterwidth{1.618cm}

\begin{tikzpicture}[-triangle 45, scale=0.65]
  \tikzstyle{dspfilter}=[rectangle, draw=black, minimum width=\filterwidth, minimum height=\filterheight, inner sep=2pt, thick]
	
	\node[] (input) at (0,0) {$x[n]$};
	\node[dspfilter] (function) at ($(input.east)+ (\horizontalsep,0)$) {$f(\cdot)$};    
	\node[dspfilter] (filter) at ($(function.east)+ (\horizontalsep,0)$) {$H(\cdot)$};  
	\node[] (output) at ($(filter)+ (\horizontalsep,0)$) {$y[n]$};
	
  
	\draw[-triangle 45] (input) -- (function);
	\draw[] (function) -- (filter);
	\draw[] (filter) -- (output);	
	
\end{tikzpicture}

%% file: fig/hammerstein_wiener.tikz
\def\horizontalsep{3cm}
\def\vertsep{7cm}

\def\filterheight{1cm}
\def\filterwidth{1.618cm}

\begin{tikzpicture}[-triangle 45, scale=0.65]
  \tikzstyle{dspfilter}=[rectangle, draw=black, minimum width=\filterwidth, minimum height=\filterheight, inner sep=2pt, thick]
	
	\node[] (input) at (0,0) {$x[n]$};
	\node[dspfilter] (function1) at ($(input.east)+ (\horizontalsep,0)$) {$f(\cdot)$};    
	\node[dspfilter] (filter1) at ($(function1.east)+ (\horizontalsep,0)$) {$H(\cdot)$};  
	\node[dspfilter] (function2) at ($(filter1.east)+ (\horizontalsep,0)$) {$g(\cdot)$};    
	\node[] (output) at ($(function2)+ (\horizontalsep,0)$) {$y[n]$};
	
  
	\draw[-triangle 45] (input) -- (function1);
	\draw[] (function1) -- (filter1);
	\draw[] (filter1) -- (function2);	
	\draw[] (function2) -- (output);	
	
\end{tikzpicture}

%% file: fig/hammerstein_parallel.tikz
\def\horizontalsep{3cm}
\def\vertsep{3cm}

\def\filterheight{1cm}
\def\filterwidth{1.618cm}

\begin{tikzpicture}[-triangle 45, scale=0.65]
  \tikzstyle{dspfilter}=[rectangle, draw=black, minimum width=\filterwidth, minimum height=\filterheight, inner sep=2pt, thick]
	
	\node[] (input) at (0,0) {$x[n]$};
	\node[dspfilter] (function1) at ($(input.east)+ (\horizontalsep,0)$) {$f^{[1]}(\cdot)$};    
	\node[dspfilter] (function2) at ($(input.east)+ (\horizontalsep,-\vertsep)$) {$f^{[2]}(\cdot)$};    	
	\node[dspfilter] (functionP) at ($(input.east)+ (\horizontalsep,-2*\vertsep)$) {$f^{[P]}(\cdot)$};    	
	
	\node[dspfilter] (filter1) at ($(function1.east)+ (\horizontalsep,0)$) {$H^{[1]}(\cdot)$};  
	\node[dspfilter] (filter2) at ($(function2.east)+ (\horizontalsep,0)$) {$H^{[2]}(\cdot)$};  	
	\node[dspfilter] (filterP) at ($(functionP.east)+ (\horizontalsep,0)$) {$H^{[P]}(\cdot)$};  
		
	\node[circle, draw=black,inner sep=0pt, thick] (sum1) at ($(filter1)+ (\horizontalsep,0)$) {\Large $\bm{+}$};		
	\node[circle, draw=black,inner sep=0pt, thick] (sum2) at ($(filter2)+ (\horizontalsep,0)$) {\Large $\bm{+}$};		
		
	\node[] (output) at ($(sum1)+ (\horizontalsep,0)$) {$y[n]$};
	
  
	\draw[-triangle 45] (input) -- (function1);
		
	\node[thick, circle, fill=black, minimum width=0.175cm, inner sep=0] at ($(input)+ (0.5*\horizontalsep,0)$) {};			
	
	\draw[-triangle 45] (input) -- ($(input)+ (0.5*\horizontalsep,0)$) |- (function2);	
	\node[thick, circle, fill=black, minimum width=0.175cm, inner sep=0] at ($(input)+ (0.5*\horizontalsep,-\vertsep)$) {};				
	
	\draw[-] (input) -| ($(input)+ (0.5*\horizontalsep,-1.25*\vertsep)$);	
	\node[] at ($(input)+ (0.5*\horizontalsep,-1.45*\vertsep)$) {$\vdots$};		
	\draw[-triangle 45] ($(input)+ (0.5*\horizontalsep,-1.75*\vertsep)$) |- (functionP);

	\draw[] (function1) -- (filter1);
	\draw[] (function2) -- (filter2);
	\draw[] (functionP) -- (filterP);
	
	\draw[] (filter1) -- (sum1);
	\draw[] (filter2) -- (sum2);
	\draw[] (filterP) -| ($(filterP) + (\horizontalsep,0.25*\vertsep)$);	
	\node[] at ($(filterP)+ (\horizontalsep,0.45*\vertsep)$) {$\vdots$};			
	\draw[] ($(filterP)+ (\horizontalsep,0.55*\vertsep)$) -- (sum2);
	
	\draw[] (sum2) -- (sum1);	
	\draw[] (sum1) -- (output);	
	
\end{tikzpicture}

%% file: fig/system.tikz
\begin{circuitikz}[scale=0.65]

	




	\draw (4,0) node[mixer,scale=0.6] (txmixer) {};
	\draw (8.5,0) node[antenna,scale=0.6] (txantenna) {};
	
	\draw (0,0) to[short,*-] ++(0,0) to[twoport,>,t=DAC] (txmixer.west) node[inputarrow]{};
	\draw (txmixer.east) to[short,-] ++(1.55,0) to[amp,>,t=\small{PA}] ++(1.5,0) to[short,-] (txantenna);
	
	\draw (0.5,0) node[above] {\small $x[n]$};
	\draw (txmixer)+(0,0.5) node[above] {\small{IQ Mixer}};
	\draw (txmixer)+(1.2,0) node[above] {\small $x_{\text{IQ}}[n]$};
	\draw (txantenna)+(-0.65,0) node[above] {\small $x_{\text{PA}}[n]$};
	
	\draw (8.5,-3) node[antenna,scale=0.6] (rxantenna) {};
	\draw (4,-3) node[mixer,scale=0.6] (rxmixer) {};
	
	\draw (rxantenna) to[short,-] ++(-1.,0) to[amp,>,t={\rotatebox[origin=c]{180}{\small{LNA}}}] ++(-1.5,0) to[short,-] (rxmixer.east) node[inputarrow,rotate=180]{};
	\draw (rxmixer.west) to[twoport,>,t=ADC] (0,-3) to[short,-*] (0,-3);
	\draw [->] (txantenna)+(0.4,1.3) to[thick, out=-20, in=20, edge node={node [right] {$h_{\text{SI}}$}}]  ($(rxantenna) + (0.4,1.35)$);
	
	\draw (0.5,-3) node[above] {\small $y[n]$};
	
	\draw (4.375,-1.5) node[oscillator,scale=0.5] (ref) {};
	\draw (ref.south) to[short,-] (rxmixer.north) node[inputarrow,rotate=270]{};
	\draw (ref.north) to[short,-] (txmixer.south) node[inputarrow,rotate=90]{};
	\draw (ref)+(-0.6,0) node[left,text width=1.2cm,align=center] {\small{Local\\Oscillator}};
	
\end{circuitikz}

%% file: fig/forward_backward.tikz
\def\horizontalbigsep{3cm}
\def\vertbigsep{2.9cm}

\def\horizontalsmallsep{3.5cm}
\def\vertsmallsep{2cm}

\def\bpsep{0.4cm}
\def\bpsepdiag{0.283cm} 
\def\bpgap{0.1cm} 

\def\filterheight{0.9cm}
\def\filterwidth{1.456cm}

\begin{tikzpicture}[-triangle 45, scale=0.65, thick]
	\tikzstyle{dspfilter}=[rectangle, draw=black, minimum width=\filterwidth, minimum height=\filterheight, inner sep=2pt, thick]
	\tikzstyle{arithmetic}=[circle, draw=black, minimum size=15pt, inner sep=-5pt, outer sep=0pt, thick]	
	
	
	\node[] (xn) at (0,0) {$x[n]$};
	\node[] (xnconj) at ($(xn) + (0,-\vertbigsep)$) {$\overline{x}[n]$};	
	
	\node[] (k1) at ($(xn) + (\horizontalsmallsep,\vertsmallsep)$) {$K_1$};
	\node[] (k2) at ($(xnconj) + (\horizontalsmallsep,\vertsmallsep)$) {$K_2$};	
		
	\node[arithmetic] (mult_xn_k1) at ($(xn) + (\horizontalsmallsep,0)$) {\Large $\bm{\times}$};	
	
	\node[arithmetic] (mult_xnconj_k2) at ($(xnconj) + (\horizontalsmallsep,0)$) {\Large $\bm{\times}$};	

	\node[arithmetic] (add_k1_k2) at ($(mult_xn_k1)!0.5!(mult_xnconj_k2) + (\horizontalsmallsep,0)$) {\Large $\bm{+}$};	
	
	\coordinate[] (xiqn) at ($(add_k1_k2) + (0.7*\horizontalsmallsep,0)$) {};		
	\coordinate[] (xiqn_bp) at ($(add_k1_k2) + (0.7*\horizontalsmallsep,0) + (\bpsepdiag, -\bpsepdiag)$) {};			
	\node[] () at ($(add_k1_k2) + (1.5,0.5)$) {$x_{\text{IQ}}[n]$};		
	
	\node[dspfilter] (function1) at ($(xiqn) + (\horizontalsmallsep,\vertbigsep)$) {$f^{[1]}(\cdot)$};    
	\node[dspfilter] (function3) at ($(xiqn) + (\horizontalsmallsep,0)$) {$f^{[3]}(\cdot)$};    	
	\node[dspfilter] (function5) at ($(xiqn) + (\horizontalsmallsep,-\vertbigsep)$) {$f^{[5]}(\cdot)$};    	
	
	\node[] (h1) at ($(function1) + (\horizontalsmallsep,\vertsmallsep)$) {$h_1[n]$};	
	\node[] (h3) at ($(function3) + (\horizontalsmallsep,\vertsmallsep)$) {$h_3[n]$};	
	\node[] (h5) at ($(function5) + (\horizontalsmallsep,\vertsmallsep)$) {$h_5[n]$};		
	
	\node[arithmetic] (mult_f1_h1) at ($(function1) + (\horizontalsmallsep,0)$) {\Large $\bm{\times}$};	
	
	\node[arithmetic] (mult_f3_h3) at ($(function3) + (\horizontalsmallsep,0)$) {\Large $\bm{\times}$};	
	
	\node[arithmetic] (mult_f5_h5) at ($(function5) + (\horizontalsmallsep,0)$) {\Large $\bm{\times}$};	

	\node[arithmetic] (mult_f5_h5) at ($(function5) + (\horizontalsmallsep,0)$) {\Large $\bm{\times}$};	

	\node[arithmetic] (add_tn) at ($(mult_f1_h1)!0.5!(mult_f5_h5) + (\horizontalsmallsep,0)$) {\Large $\bm{+}$};

	\node[dspfilter] (sub_y_t) at ($(add_tn) + (\horizontalsmallsep,0)$) {$\text{MSE}(\cdot)$};    	
	\node[] (target) at ($(sub_y_t.north) + (0,\vertsmallsep) + (0, -10pt)$) {$t[n]$};					
	\node[] () at ($(add_tn) + (1.5,0.5)$) {$y[n]$};		

	\node[] (cn) at ($(sub_y_t) + (\horizontalsmallsep,0)$) {$C[n]$};


	\draw[blue, densely dashed, densely dashed] ($(sub_y_t.west) + (-\bpgap,-\bpsep)$) -- node [below, pos=0.4] {$\dfrac{\partial C}{\partial \overline{y}[n]}$} ($(add_tn.east) + (0,-\bpsep)$);

	\draw[blue, densely dashed] ($(add_tn.north) + (0,\bpgap)$) -- ($(mult_f1_h1.east) + (\bpgap,0)$);
	\draw[blue, densely dashed] ($(add_tn.west) + (-\bpgap,-\bpsepdiag)$) -- ($(mult_f3_h3.east) + (\bpgap,-\bpsepdiag)$);	
	\draw[blue, densely dashed] ($(add_tn.south) + (0,-\bpgap)$) -- ($(mult_f5_h5.east) + (\bpgap,0)$);

	\draw[blue, densely dashed] ($(mult_f1_h1) + (-\bpsepdiag,\bpsepdiag) + (0, 2*\bpgap)$) -- node [left, pos=1.0] {$\dfrac{\partial C}{\partial \overline{h}_1[n]}$} ($(h1.south) + (-\bpsepdiag,0)$);
	\draw[blue, densely dashed] ($(mult_f3_h3) + (-\bpsepdiag,\bpsepdiag) + (0, 2*\bpgap)$) -- node [left, pos=1.0] {$\dfrac{\partial C}{\partial \overline{h}_3[n]}$} ($(h3.south) + (-\bpsepdiag,0)$);
	\draw[blue, densely dashed] ($(mult_f5_h5) + (-\bpsepdiag,\bpsepdiag) + (0, 2*\bpgap)$) -- node [left, pos=1.0] {$\dfrac{\partial C}{\partial \overline{h}_5[n]}$} ($(h5.south) + (-\bpsepdiag,0)$);

	\draw[blue, densely dashed] ($(mult_f1_h1.west) + (-\bpgap,-\bpsepdiag)$) -- ($(function1.east) + (\bpgap,-\bpsepdiag)$);	
	\draw[blue, densely dashed] ($(mult_f3_h3.west) + (-\bpgap,-\bpsepdiag)$) -- ($(function3.east) + (\bpgap,-\bpsepdiag)$);	
	\draw[blue, densely dashed] ($(mult_f5_h5.west) + (-\bpgap,-\bpsepdiag)$) -- ($(function5.east) + (\bpgap,-\bpsepdiag)$);	
	
	\draw[blue, densely dashed, -] ($(function1.west) + (-\bpgap,-\bpsepdiag)$) -- ($(xiqn_bp |- function1) + (0,-\bpsepdiag)$) -- ($(xiqn_bp) + (0,0)$);	
	\draw[blue, densely dashed, -] ($(function3.west) + (-\bpgap,-\bpsepdiag)$) -- ($(xiqn_bp) + (0,0)$);	
	\draw[blue, densely dashed, -] ($(function5.west) + (-\bpgap,-\bpsepdiag)$) -- ($(xiqn_bp |- function5) + (0,-\bpsepdiag)$) -- ($(xiqn_bp) + (0,0)$);	
	\node[thick, circle, fill=blue, minimum width=0.175cm, inner sep=0] at (xiqn_bp) {};		
	
	\draw[blue, densely dashed] (xiqn_bp) -- node [below, pos=0.6] {$\dfrac{\partial C}{\partial \overline{x}_{\text{IQ}}[n]}$} ($(add_k1_k2.east) + (\bpgap,-\bpsepdiag)$);

	\draw[blue, densely dashed] ($(add_k1_k2.north) + (-0,\bpgap)$) -- ($(mult_xn_k1.east) + (\bpgap,\bpsepdiag)$);
	\draw[blue, densely dashed] ($(add_k1_k2.south) + (-0,-\bpgap)$) -- ($(mult_xnconj_k2.east) + (\bpgap,-\bpsepdiag)$);

	\draw[blue, densely dashed] ($(mult_xn_k1) + (-\bpsepdiag,\bpsepdiag) + (0, 2*\bpgap)$) -- node [left, pos=1.0] {$\dfrac{\partial C}{\partial \overline{K}_1}$} ($(k1.south) + (-\bpsepdiag,0)$);
	\draw[blue, densely dashed] ($(mult_xnconj_k2) + (-\bpsepdiag,\bpsepdiag) + (0, 2*\bpgap)$) -- node [left, pos=1.0] {$\dfrac{\partial C}{\partial \overline{K}_2}$} ($(k2.south) + (-\bpsepdiag,0)$);

	
	\draw (xn) -- (mult_xn_k1);
	\draw (k1) -- (mult_xn_k1);		
		
	\draw (xnconj) -- (mult_xnconj_k2);
	\draw (k2) -- (mult_xnconj_k2);			
		
	\draw (mult_xn_k1) -- (add_k1_k2);	 
	\draw (mult_xnconj_k2) -- (add_k1_k2);			
		
	\draw[-] (add_k1_k2) -- (xiqn);						
	\node[thick, circle, fill=black, minimum width=0.175cm, inner sep=0] at (xiqn) {};		
		
	\draw (add_k1_k2) -- (xiqn) |- (function1);				
	\draw (add_k1_k2) -- (xiqn) -- (function3);						
	\draw (add_k1_k2) -- (xiqn) |- (function5);						
		
	\draw (h1) -- (mult_f1_h1);
	\draw (function1) -- (mult_f1_h1);	
	
	\draw (h3) -- (mult_f3_h3);
	\draw (function3) -- (mult_f3_h3);	
	
	\draw (h5) -- (mult_f5_h5);		
	\draw (function5) -- (mult_f5_h5);				
		
	\draw (mult_f1_h1) -- (add_tn);	 
	\draw (mult_f3_h3) -- (add_tn);			
	\draw (mult_f5_h5) -- (add_tn);					
		
	\draw (target) -- (sub_y_t);			
	\draw (add_tn) -- (sub_y_t);				
		
	\draw (sub_y_t) -- (cn);						
			
	
\end{tikzpicture}

%% file: fig/train_plot_hammerstein.tikz
\begin{tikzpicture}
    \begin{axis}[
		normalsize,
		width = 7.6cm,
		height = 4.7cm,
		xmin=1, xmax=50,
		ymin=36, ymax=45.5,
		ymajorgrids=true,
		grid style=dashed,
		xlabel = {Epoch},
		ylabel = {SI Cancellation (dB)},
		ylabel near ticks,
		xlabel near ticks,
		xtick distance=5,
		ytick distance=2,
		label style={font=\small},
		tick label style={font=\small},
		ymajorgrids,
		legend pos = south east,
		legend style={font=\footnotesize},
		legend columns=1,
		legend cell align=left,
    ]

	\addplot[black, very thick, solid] coordinates {(0,44.5) (50,44.5)};
	\addlegendentry{WLMP}

	\addplot[Set1-7-1, very thick, solid] table[x index = 0, y index = 1] {fig/results_ftrl/pgf_dat_train/hammerstein_all_struct_C-C_min_power_1_max_power_5_test_cancellation_n_epochs_50_train_size_1_0_batch_size_6_lr_0_2628534593844867_initializer_hammerstein_rayleigh.dat};
    \addlegendentry{Model-based NN}
	\addplot+[name path=test_htnnbot, color=Set1-7-1, mark=none, forget plot] table[x index = 0, y index = 2] {fig/results_ftrl/pgf_dat_train/hammerstein_all_struct_C-C_min_power_1_max_power_5_test_cancellation_n_epochs_50_train_size_1_0_batch_size_6_lr_0_2628534593844867_initializer_hammerstein_rayleigh.dat};
	\addplot+[name path=test_htnntop, color=Set1-7-1, mark=none, forget plot] table[x index = 0, y index = 3] {fig/results_ftrl/pgf_dat_train/hammerstein_all_struct_C-C_min_power_1_max_power_5_test_cancellation_n_epochs_50_train_size_1_0_batch_size_6_lr_0_2628534593844867_initializer_hammerstein_rayleigh.dat};
	\addplot[Set1-7-1!50,fill opacity=0.5, forget plot] fill between[of=test_htnnbot and test_htnntop];

	\addplot[Set1-7-3, very thick, solid] table[x expr=\thisrow{i}, y expr=\thisrow{x}+37.85998344421386] {fig/results_adam/pgf_dat_train/complex_ffnn_nl_struct_6_test_cancellation_n_epochs_50_train_size_1_0_batch_size_39_lr_0_003887654113612277.dat};
    \addlegendentry{CVNN}
	\addplot+[name path=test_cvnnbot, color=Set1-7-3, mark=none, forget plot] table[x expr=\thisrow{i}, y expr=\thisrow{y}+37.85998344421386] {fig/results_adam/pgf_dat_train/complex_ffnn_nl_struct_6_test_cancellation_n_epochs_50_train_size_1_0_batch_size_39_lr_0_003887654113612277.dat};
	\addplot+[name path=test_cvnntop, color=Set1-7-3, mark=none, forget plot] table[x expr=\thisrow{i}, y expr=\thisrow{z}+37.85998344421386] {fig/results_adam/pgf_dat_train/complex_ffnn_nl_struct_6_test_cancellation_n_epochs_50_train_size_1_0_batch_size_39_lr_0_003887654113612277.dat};
	\addplot[Set1-7-3!50,fill opacity=0.5, forget plot] fill between[of=test_cvnnbot and test_cvnntop];

    \end{axis}
\end{tikzpicture}

%% file: fig/complexity_group_plot.tikz
\begin{tikzpicture}

	\pgfplotsset{grid style={dashed}}

	\begin{groupplot}[
		group style={
			group size=2 by 1,
			xlabels at=edge bottom,
			xticklabels at=edge bottom,
			horizontal sep=0.6cm,
			vertical sep=0pt.
	    },
		ytick distance=1,
		label style={font=\small},
		tick label style={font=\footnotesize},
		yticklabel style={
						/pgf/number format/fixed,
						/pgf/number format/precision=0,
						/pgf/number format/zerofill
		},
		width = 5cm,
		height = 4.1cm,
		ymajorgrids,
		legend style={at={(0.3,-0.52)},anchor=west},
		legend style={font=\scriptsize},
		legend columns=3,
		legend cell align=left,
	]
	\nextgroupplot[xmin = 25, xmax = 450, ymin = 39, ymax = 46, xmode=log, log ticks with fixed point, xlabel = {\# Real-Valued Parameters}, ylabel={SI Cancellation (dB)}, y label/.style={rotate=-90, xshift=.5cm}]
	\addplot[black, thick, solid, mark=x] table[x index = 0, y index = 1] {fig/results_poly/pgf_dat_complexity/mem_polynomial_nl_complexity_mult_algo_reduced_cmult.dat};
	\addlegendentry{WLMP};
	\addplot[Set1-7-1, thick, solid, mark=diamond*] table[x index = 0, y index = 1] {fig/results_ftrl/pgf_dat_complexity/mem_hammerstein_all_complexity_mult_algo_reduced_cmult_cc.dat};
	\addlegendentry{Model-based NN};
	\addplot[Set1-7-3, thick, solid, mark=triangle*] table[x index = 0, y index = 1] {fig/results_adam/pgf_dat_complexity/mem_complex_ffnn_nl_complexity_mult_algo_reduced_cmult_shallow.dat};
	\addlegendentry{CVNN};

	\nextgroupplot[xmin = 250, xmax = 4500, ymin = 39, ymax = 46, xmode=log, log ticks with fixed point, xlabel = {\# Real-Valued FLOPs}]
	\addplot[black, thick, solid, mark=x] table[x index = 0, y index = 1] {fig/results_poly/pgf_dat_complexity/flop_polynomial_nl_complexity_mult_algo_reduced_cmult.dat};
	\addplot[Set1-7-1, thick, solid, mark=diamond*] table[x index = 0, y index = 1] {fig/results_ftrl/pgf_dat_complexity/flop_hammerstein_all_complexity_mult_algo_reduced_cmult_cc.dat};
	\addplot[Set1-7-3, thick, solid, mark=triangle*] table[x index = 0, y index = 1] {fig/results_adam/pgf_dat_complexity/flop_complex_ffnn_nl_complexity_mult_algo_reduced_cmult_shallow.dat};
	\end{groupplot}
\end{tikzpicture}